\documentclass[12pt]{article}
\usepackage{amsmath}
\usepackage{graphicx}
\usepackage{subfigure}
\usepackage{hyperref}

\begin{document}

\author{Ernst Trojan and George V. Vlasov \and \textit{Moscow Institute of Physics
and Technology} \and \textit{PO Box 3, Moscow, 125080, Russia}}
\title{Thermodynamics of exotic matter with constant $w=P/E$}
\maketitle

\begin{abstract}
We consider a substance with equation of state $P=wE$ at constant $w$ and
find that it is an ideal gas of quasi-particles with the energy spectrum $%
\varepsilon _p\sim p^{wq}$ that can constitute either regular matter (when $%
w>0$) or exotic matter (when $w<0$) in a $q$-dimensional space.
Particularly, an ideal gas of fermions or bosons with the energy spectrum $%
\varepsilon _p=m^4/p^3$ in 3-dimensional space will have the pressure $P=-E$%
. Exotic material, associated with the dark energy at $E+P<0$, is also
included in analysis. We determine the properties of regular and exotic
ideal Fermi gas at zero temperature and derive a low-temperature expansion
of its thermodynamical functions at finite temperature. The Fermi level of
exotic matter is shifted below the Fermi energy at zero temperature, while
the Fermi level of regular matter is always above it. The heat capacity of
any fermionic substance is always linear dependent on temperature, but
exotic matter has negative entropy and negative heat capacity.
\end{abstract}

\section{Introduction}

The equation of state (EOS) is a fundamental characteristic of matter. It is
a functional link $P\left( E\right) $\ between the pressure $P$\ and the
energy density $E$. Its knowledge allows to predict the behavior of
substance which appears in various problems in astrophysics, including
cosmology and physics of neutron stars.

The EOS can be given by expression 
\begin{equation}
P=wE  \label{xi}
\end{equation}
where $w$ is a dimensionless parameter that, in general, is a function
depending on $E$. Particularly, an ideal gas of non-relativistic particles
has constant $w=2/3$. The EOS with $w=1/3$ describes radiation and phonon
gas, while the EOS of dust has $w=0$. One of most exotic examples of EOS 
\begin{equation}
P=E  \label{sti}
\end{equation}
corresponds to the so-called 'absolute stiff' matter, that may appear in
various problems of astrophysics.

Most forms of matter available for experimental research exist at $0<w<1$
and have positive pressure and positive energy density. Other forms of
substance is commonly called as exotic matter. It often appears in applied
problems of astrophysics.\textrm{\ }Particularly, the tachyon matter can
have $w>1$ \cite{TV2011c,TV2011d}, while materials with negative $w<0$ are
considered in cosmology as candidates for the dark energy. Researches do not
stop their efforts for constructing the EOS of such exotic substances \cite
{dark1,dark2,dark3}. Of course, it highly desirable to have a ready-made
model for calculating the thermodynamical parameters of exotic matter.
However, it is still uncertain what physical particles could form this
material. It is clear that neither free massive particles with the energy
spectrum 
\begin{equation}
\varepsilon _p=\sqrt{p^2+m^2}  \label{pro}
\end{equation}
nor tachyons with the energy spectrum 
\begin{equation}
\varepsilon _p=\sqrt{m^2-p^2}  \label{pro2}
\end{equation}
could yield negative $w$ in the EOS $P=wE$. However, the interaction between
particles can be responsible for exotic forms of the EOS. For example, the
dense nuclear matter in the interiors of neutron stars has almost 'absolute
stiff' EOS (\ref{sti}).

It is clear that other exotic forms of EOS also belong to a strongly
interacting medium, and its further description is not possible without
solving the quantum many-body problem. Nevertheless, a system of real
interacting particles can be modeled by a system of free hypothetical
particles moving in some external field \cite{DG90}. For example, the EOS of
'absolute stiff' matter can be modeled by an ideal gas of free particles
with the energy spectrum $\varepsilon _p=p^3/m^2$ \cite{TV2011e}.\ Of
course, such hypothetical particles, better to say, quais-particles, do not
exist in nature, and it is no more than a model for description of strongly
interacting medium.

There is principal restriction to apply this model of free quasi-particles
for description of substances that appear in various astrophysical problems.
Such substance can be regular matter ($E>0$, $P>0$), or exotic matter with
positive energy $E>0$ and negative pressure $P<0$, as well as exotic matter
with negative energy $E<0$ and positive pressure $P>0$. Particularly, the
phantom matter with $E+P<0$ attracts special interest.

In the present paper we consider exotic matter with the EOS $P=wE$ (\ref{xi}%
) at constant $w$. We know nothing about its thermodynamical functions and
we need to establish the energy spectrum of quasi-particles that can
constitute this substance when it is regular matter (at $w>0$) or exotic
matter (at $w<0$). Then, we can study the properties of regular and exotic
Fermi gas at zero temperature and derive the low temperature expansion of
its thermodynamical functions at finite temperature. It is also important to
a low-temperature behavior of the Fermi level and the heat capacity of
fermionic exotic matter.

Standard relativistic units $c_{light}=\hbar =k_B=1$\ are used in the paper.

\section{Thermodynamical functions}

Consider an ideal gas of free particles with the single-particle energy
spectrum $\varepsilon _p$ at finite temperature $T$ and in a $q$-dimensional
space. Let $\mu $ be the chemical potential of this system. The particle
number density $n$, pressure $P$ and energy density $E$ are determined by
standard formulas \cite{Kapusta89} 
\begin{equation}
n=\frac \gamma {\left( 2\pi \right) ^q}\int\limits_0^\infty f_p\,d^qp
\label{n}
\end{equation}
\begin{equation}
P=-T\ln Z  \label{p}
\end{equation}
\begin{equation}
E=\frac \gamma {\left( 2\pi \right) ^q}\int\limits_0^\infty f_p\varepsilon
_pd^qp  \label{ene}
\end{equation}
where 
\begin{equation}
f_p=\frac 1{\exp \left[ (\varepsilon _p-\mu )/T\right] \pm 1}  \label{f}
\end{equation}
is the distribution function, while 
\begin{equation}
\ln Z=\mp \frac \gamma {\left( 2\pi \right) ^q}\int\limits_0^\infty \ln
\left\{ 1\pm \exp \left[ (\varepsilon _p-\mu )/T\right] \right\} d^qp
\label{z}
\end{equation}
is the statistical sum, and the sign ''$+$''or ''$-$'' corresponds to
fermions and bosons. The volume of $q$-dimensional hypersphere is defined as 
\begin{equation}
d^qp=\frac{q\pi ^{q/2}}{\Gamma \left( \frac q2+1\right) }p^{q-1}dp
\label{imp}
\end{equation}
Partial integration of (\ref{z}) and its substitution in (\ref{z}) yields 
\begin{equation}
P=\frac \gamma {\left( 2\pi \right) ^q}\frac{\pi ^{q/2}}{\Gamma \left( \frac
q2+1\right) }\int\limits_0^\infty f_p\frac{\partial \varepsilon _p}{\partial
p}p^qdp  \label{p0}
\end{equation}
For example, in $3$ dimensions 
\begin{equation}
d^3p=4\pi p^2dp  \label{imp3}
\end{equation}
and 
\begin{equation}
P=\frac \gamma {6\pi ^2}\int\limits_0^\infty f_p\frac{\partial \varepsilon _p%
}{\partial p}p^3dp  \label{p03}
\end{equation}

Let us imagine that the medium with equation of state $P=wE$ is an ideal gas
of free quasi-particles with the energy spectrum $\varepsilon _p$. From (\ref
{xi}), (\ref{ene}) and (\ref{p0}) we get equation: 
\begin{equation}
P-wE=\frac \gamma {\left( 2\pi \right) ^q}\frac{q\pi ^{q/2}}{\Gamma \left(
\frac q2+1\right) }\int\limits_0^\infty f_p\left( \frac pq\frac{d\varepsilon
_p}{dp}-w\varepsilon _p\right) p^{q-1}dp=0  \label{zer}
\end{equation}
whose solution is 
\begin{equation}
\varepsilon _p=ap^{wq}  \label{e0}
\end{equation}
where $a$ is an arbitrary constant which can be either positive or negative.
Expression (\ref{e0}) differs from the standard single-particle energy
spectrum of free particles (\ref{pro}) and the objects with the energy
spectrum (\ref{e0}) should be referred as excitations or quasi-particles. In
a 3-dimensional space the energy spectrum 
\begin{equation}
\varepsilon _p=\frac{p^3}{M^2}  \label{e1}
\end{equation}
belongs to the 'absolute stiff' matter (\ref{sti}) \cite{TV2011e}, the dust
material has the energy spectrum $\varepsilon _p=M=\mathrm{const}$ that
corresponds to $w=0$, while the exotic matter with $P=-E$ is composed of
quasi-particles with energy 
\begin{equation}
\tilde \varepsilon _p=\frac{M^4}{p^3}  \label{e-1}
\end{equation}
where parameter $M$ has dimension of mass. The same EOS\ $P=-E$ is obtained
with quasi-particles whose energy is negative 
\begin{equation}
\tilde \varepsilon _p=-\frac{M^4}{p^3}  \label{e-1z}
\end{equation}

Let us introduce dimensionless variable 
\begin{equation}
x=\frac{\left| a\right| p^{wq}}T  \label{z00}
\end{equation}
and 
\begin{equation}
\Sigma _q=\frac \gamma {\left( 2\pi \right) ^q}\frac{\pi ^{q/2}}{\Gamma
\left( \frac q2+1\right) }  \label{z1}
\end{equation}
The energy spectrum (\ref{e0}) and the distribution function (\ref{f}) will
be presented so 
\begin{equation}
\varepsilon _p=xT\mathrm{sign}\left( a\right)   \label{e000}
\end{equation}
and 
\begin{equation}
f_p\left( x\right) =\frac 1{\exp \left[ \mathrm{sign}\left( a\right) x-\mu
/T\right] \pm 1}  \label{fall1}
\end{equation}
where $\mathrm{sign}\left( a\right) =1$ for positive $a>0$, and $\mathrm{sign%
}\left( a\right) =-1$ for negative $a<0$.

Limits of integration in (\ref{n}) and (\ref{ene}) correspond to 
\begin{equation}
x\left( 0\right) =\frac{\left| a\right| }T\underset{p\rightarrow 0}{\lim }%
\,p^{wq}\qquad x\left( \infty \right) =\frac{\left| a\right| }T\underset{%
p\rightarrow \infty }{\lim }\,p^{wq}  \label{plim0}
\end{equation}
that at positive $w>0$ implies 
\begin{equation}
x\left( 0\right) =0\qquad x\left( \infty \right) =\infty   \label{plim}
\end{equation}
while at negative $w<0$ expression (\ref{plim0}) implies 
\begin{equation}
x\left( 0\right) =\infty \qquad x\left( \infty \right) =0  \label{plim2}
\end{equation}
Then, substituting (\ref{e000}) together with (\ref{z00})-(\ref{z1}) in (\ref
{n}) and (\ref{ene}), we determine universal formulas for particle number
density 
\begin{equation}
n=\frac{\Sigma _q}w\frac{T^{1/w}}{\left| a\right| ^{1/w}}\int\limits_{x%
\left( 0\right) }^{x\left( \infty \right) }\frac{x^{1/w-1}dx}{\exp \left[ 
\mathrm{sign}\left( a\right) x-\mu /T\right] \pm 1}  \label{nn}
\end{equation}
and the energy density 
\begin{equation}
E=\mathrm{sign}\left( a\right) \frac{\Sigma _q}w\frac{T^{1/w+1}}{\left|
a\right| ^{1/w}}\int\limits_{x\left( 0\right) }^{x\left( \infty \right) }%
\frac{x^{1/w}dx}{\exp \left[ \mathrm{sign}\left( a\right) \left( x-\mu
/T\right) \right] \pm 1}  \label{ee}
\end{equation}
corresponding to the EOS $P=wE$ with constant $w$. 

\begin{figure}[tbp]
\caption{Fermi distribution function at $a>0$.}
\label{exo1}{\includegraphics[scale=0.6]{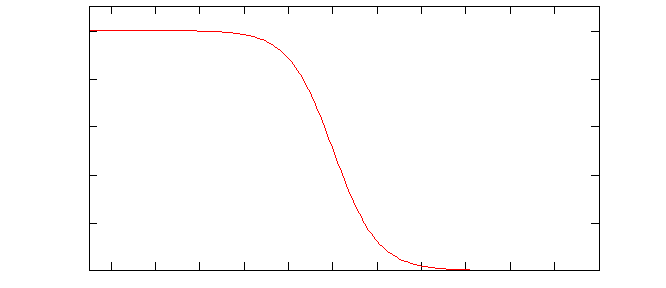}}
\end{figure}

\begin{figure}[tbp]
\caption{Fermi distribution function at $a<0$.}
\label{exo2}{\includegraphics[scale=0.6]{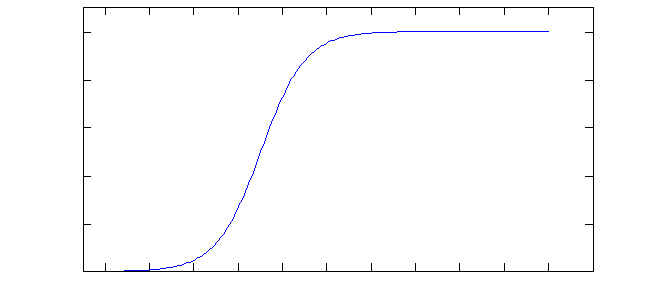}}
\end{figure}

According to formulas (\ref{nn}) and (\ref{ee}), we can calculate
thermodynamical functions of a Fermi gas at low temperature. At $a>0$ the
chemical potential is positive $\mu >0$, and the distribution function (\ref
{fall1}) is plotted in Fig.~\ref{exo1}. At $a<0$ the chemical potential is
negative $\mu <0$, and the distribution function (\ref{fall1}) is given in
Fig.~\ref{exo2}. So, the distribution function of a Fermi gas can be
presented in the universal form 
\begin{equation}
f_p\left( x,\lambda \right) =\frac 1{\exp \left[ \mathrm{sign}\left(
a\right) \left( x-\lambda \right) \right] +1}  \label{ff}
\end{equation}
where 
\begin{equation}
\lambda =\frac{\left| \mu \right| }T  \label{lam}
\end{equation}
Then, taking into account (\ref{plim}) and (\ref{plim2}) we can rewrite (\ref
{nn}) and (\ref{ee}) in the form 
\begin{equation}
n=\frac{\Sigma _q}{\left| w\right| }\frac{T^{1/w}}{\left| a\right| ^{1/w}}%
\int\limits_0^\infty \frac{x^{1/w-1}dx}{\exp \left[ \mathrm{sign}\left(
a\right) \left( x-\lambda \right) \right] +1}  \label{nno}
\end{equation}
\begin{equation}
E=\mathrm{sign}\left( a\right) \frac{\Sigma _q}{\left| w\right| }\frac{%
T^{1/w+1}}{\left| a\right| ^{1/w}}\int\limits_0^\infty \frac{x^{1/w+1}dx}{%
\exp \left[ \mathrm{sign}\left( a\right) \left( x-\lambda \right) \right] +1}
\label{eeo}
\end{equation}

The EOS\ $P=wE$ (\ref{xi}) imposes no restriction concerning the signs of $a$
and $w$ in the energy spectrum (\ref{e0}). The regular matter is
characterized by 
\begin{equation}
a>0\qquad w>0\qquad \Leftrightarrow \qquad E>0\qquad P>0  \label{reg}
\end{equation}
while negative $a<0$ and positive $w>0$ corresponds to the material with
negative pressure and negative energy density: 
\begin{equation}
a<0\qquad w>0\qquad \Leftrightarrow \qquad E<0\qquad P<0  \label{al3}
\end{equation}
Our main interest is focused on the exotic matter that has negative $w<0$
and whose energy spectrum (\ref{e0}) admits two alternatives

\begin{equation}
a>0\qquad w<0\qquad \Leftrightarrow \qquad E>0\qquad P<0  \label{al1}
\end{equation}
and 
\begin{equation}
a<0\qquad w<0\qquad \Leftrightarrow \qquad E<0\qquad P>0  \label{al2}
\end{equation}

\section{Exotic fermion matter at zero temperature}

Consider an ideal Fermi gas whose EOS is $P=wE$. This gas is composed of
quasi-particles with the energy spectrum (\ref{e0}). The distribution
function  of a Fermi gas (\ref{ff}) at low temperature (large $\lambda \gg 1$%
) reveals the following asymptotic behavior 
\begin{equation}
f_p\left( 0,\lambda \right) =\frac 1{\exp \left[ -\mathrm{sign}\left(
a\right) \lambda \right] +1}\cong \Theta \left( a\right) \qquad \underset{%
x\rightarrow \infty }{\lim }f_p\left( x,\lambda \right) =\Theta \left(
-a\right)   \label{ff1}
\end{equation}
In other words 
\begin{equation}
f_p\left( 0,\lambda \right) =1\qquad \underset{x\rightarrow \infty }{\lim }%
f_p\left( x,\lambda \right) =0\qquad a>0  \label{ff2}
\end{equation}
and 
\begin{equation}
f_p\left( 0,\lambda \right) =0\qquad \underset{x\rightarrow \infty }{\lim }%
f_p\left( x,\lambda \right) =1\qquad a<0  \label{ff3}
\end{equation}
At very low temperature 
\begin{equation}
\lambda \rightarrow \frac{\left| \varepsilon _F\right| }T  \label{z000}
\end{equation}
where 
\begin{equation}
\varepsilon _F=ap_F^{wq}  \label{lev}
\end{equation}
is the Fermi energy and $p_F$ is the Fermi momentum, and the distribution
function it approaches to the Heaviside step 
\begin{equation}
f_p\rightarrow \Theta \left[ \mathrm{sign}\left( a\right) \left( \left|
\varepsilon _F\right| /T-x\right) \right]   \label{step}
\end{equation}
At zero temperature distribution function is taken in the form 
\begin{equation}
f_p=\Theta \left( \varepsilon _F-\varepsilon _p\right) =\Theta \left[ 
\mathrm{sign}\left( a\right) \left( \left| \varepsilon _F\right| -\left|
\varepsilon \right| \right) \right]   \label{fe1}
\end{equation}
where the energy spectrum is determined by formula (\ref{e0}).

For regular matter (\ref{reg}) the distribution function (\ref{fe1}) is
equivalent to 
\begin{equation}
f_p=\Theta \left( p_F-p\right)  \label{f1}
\end{equation}
and limits of integration $p\in \left( 0,p_F\right) $ correspond to $%
\varepsilon \in \left( 0,\varepsilon _F\right) $. At zero temperature $%
T\rightarrow 0$ formulas (\ref{nno}) and (\ref{eeo}) yield 
\begin{equation}
n=\frac{\Sigma _q}{wa^{1/w}}\underset{T\rightarrow 0}{\lim }\left(
T^{1/w}\int\limits_0^\infty x^{1/w-1}dx\right) =\frac{\Sigma _q}{wa^{1/w}}%
\int\limits_0^{\varepsilon _F}\varepsilon ^{1/w-1}d\varepsilon =\Sigma
_q\left( \frac{\varepsilon _F}a\right) ^{1/w}=\Sigma _q\,p_F^q  \label{nn1}
\end{equation}
\begin{equation}
E=\frac{\Sigma _q}{wa^{1/w}}\left( \underset{T\rightarrow 0}{\lim }%
T^{1/w+1}\int\limits_0^\infty x^{1/w}dx\right) =\frac{\Sigma _q}{wa^{1/w}}%
\int\limits_0^{\varepsilon _F}\varepsilon ^{1/w}d\varepsilon =\Sigma _q\frac{%
\varepsilon _F^{1/w+1}}{a^{1/w}}=\frac{a\Sigma _q}{w+1}p_F^{\left(
w+1\right) q}  \label{ee1}
\end{equation}
Hence, formulas (\ref{nn1}) and (\ref{ee1}) imply 
\begin{equation}
E=\frac a{\left( w+1\right) \Sigma _q^w}n^{w+1}  \label{en11}
\end{equation}
Particularly, for the 'absolute stiff' matter with $w=1$\ we have always 
\cite{TV2011e}\textrm{\ } 
\begin{equation}
P=E=\frac a{2\Sigma _q}n^2  \label{p1q}
\end{equation}

For exotic matter (\ref{e0}) with $a>0$ and $w<0$ the distribution function (%
\ref{fe1}) is equivalent to 
\begin{equation}
f_p=\Theta \left( p-p_F\right)  \label{f2}
\end{equation}
that determines limits of integration $p\in \left( p_F,\infty \right) $
corresponding to $\varepsilon \in \left( \varepsilon _F,0\right) $. At zero
temperature formulas (\ref{nno}) and (\ref{eeo}) yield 
\begin{equation}
n=-\frac{\Sigma _q}{wa^{1/w}}\underset{T\rightarrow 0}{\lim }\left(
T^{1/w}\int\limits_0^\infty x^{1/w-1}dx\right) =\frac{\Sigma _q}{wa^{1/w}}%
\int\limits_{\varepsilon _F}^0\varepsilon ^{1/w-1}d\varepsilon =q\Sigma
_q\int\limits_{p_F}^\infty p^{q-1}dp  \label{nn2}
\end{equation}
\begin{equation}
E=-\frac{\Sigma _q}{wa^{1/w}}\left( \underset{T\rightarrow 0}{\lim }%
T^{1/w+1}\int\limits_0^\infty x^{1/w}dx\right) =\frac{\Sigma _q}{wa^{1/w}}%
\int\limits_{\varepsilon _F}^0\varepsilon ^{1/w}d\varepsilon =aq\Sigma
_q\int\limits_{p_F}^\infty p^{wq}p^{q-1}dp  \label{ee2}
\end{equation}
Integral (\ref{nn2}) is divergent at $w<0$, integral (\ref{ee2}) is
divergent at $0\geq w\geq -1$. At $w<-1$ integral (\ref{ee2}) is finite and
evaluated as 
\begin{equation}
E=-\frac{\Sigma _q}{wa^{1/w}}\int\limits_0^{\varepsilon _F}\varepsilon
^{1/w}d\varepsilon =-\frac{\Sigma _q}{\left( w+1\right) a^{1/w}}\varepsilon
_F^{1/w+1}=-\frac{a\Sigma _q}{w+1}p_F^{\left( w+1\right) q}  \label{ee22}
\end{equation}
However, the particle number density (\ref{nn2}) remains undefined. If we
introduce an upper cutoff momentum $p_0\gg p_F$, integral (\ref{ee22}) is
estimated as 
\begin{equation}
n\simeq \Sigma _qp_0^q\gg \Sigma _qp_F^q  \label{nn21}
\end{equation}
but integral (\ref{ee22}) remains unchanged 
\begin{equation}
E=\frac{a\Sigma _q}{w+1}\left[ p_0^{\left( w+1\right) q}-p_F^{\left(
w+1\right) q}\right] \simeq -\frac{a\Sigma _q}{w+1}p_F^{\left( w+1\right) q}
\label{ee23}
\end{equation}
The pressure $P=wE$ is negative, while $P+E<0$ at $w<-1$.

For exotic matter with negative $a<0$ (\ref{al2}) the energy $\varepsilon _p$
(\ref{e0}) and the Fermi level $\varepsilon _F$ (\ref{f1}) are negative, and
distribution function (\ref{fe1}) and (\ref{fx1}) implies 
\begin{equation}
f_p=\Theta \left( \left| \varepsilon _p\right| -\left| \varepsilon _F\right|
\right)  \label{fe2}
\end{equation}
At negative $w<0$ this distribution function is equivalent to (\ref{f1}) but
limits of integration $p\in \left( 0,p_F\right) $ correspond to $\varepsilon
_p\in \left( -\infty ,\varepsilon _F\right) $. At zero temperature formulas (%
\ref{nno}) and (\ref{eeo}) yield 
\begin{equation}
n=-\frac{\Sigma _q}{w\left| a\right| ^{1/w}}\left( \underset{T\rightarrow 0%
}{\lim }T^{1/w}\int\limits_0^\infty x^{1/w-1}dx\right) =\frac{\Sigma _q}{%
w\left| a\right| ^{1/w}}\int\limits_\infty ^{\left| \varepsilon _F\right|
}\xi ^{1/w-1}d\xi =\Sigma _q\left( \frac{\varepsilon _F}a\right)
^{1/w}=\Sigma _q\,p_F^q  \label{nn3}
\end{equation}
\begin{equation}
E=\frac{\Sigma _q}{w\left| a\right| ^{1/w}}\left( \underset{T\rightarrow 0%
}{\lim }T^{1/w+1}\int\limits_0^\infty x^{1/w}dx\right) =-\frac{\Sigma _q}{%
w\left| a\right| ^{1/w}}\int\limits_\infty ^{\left| \varepsilon _F\right|
}\xi ^{1/w}d\xi =-\left| a\right| q\Sigma
_q\int\limits_0^{p_F}p^{wq}p^{q-1}dp  \label{ee3}
\end{equation}
Integral (\ref{ee3}) is divergent at $w\leq -1$. However, at $0>w>-1$
formulas (\ref{nn3}) and (\ref{ee3}) are fully integrated, resulting in 
\begin{equation}
E=-\frac{\left| a\right| }{\left( w+1\right) \Sigma _q^w}n^{w+1}<0
\label{p1x}
\end{equation}
that is similar to (\ref{en11}). This exotic matter has negative energy
density and positive pressure, meanwhile $P+E<0$.

For exotic matter with negative $a<0$ and positive $w>0$ (\ref{al3}) the
distribution function (\ref{fe2}) is equivalent to (\ref{f1}). Hence, limits
of integration $p\in \left( p_F,\infty \right) $ will correspond to $%
\varepsilon _p\in \left( -\varepsilon _F,-\infty \right) $. The particle
number density and the energy density are determined by formulas

\begin{equation}
n=\frac{\Sigma _q}{w\left| a\right| ^{1/w}}\left( \underset{T\rightarrow 0%
}{\lim }T^{1/w}\int\limits_0^\infty x^{1/w-1}dx\right) =\frac{\Sigma _q}{%
w\left| a\right| ^{1/w}}\int\limits_{\left| \varepsilon _F\right| }^\infty
\xi ^{1/w-1}d\xi =q\Sigma _q\int\limits_{p_F}^\infty p^{q-1}dp  \label{nn334}
\end{equation}
\begin{equation}
E=-\frac{\Sigma _q}{w\left| a\right| ^{1/w}}\left( \underset{T\rightarrow 0%
}{\lim }T^{1/w+1}\int\limits_0^\infty x^{1/w}dx\right) =-\frac{\Sigma _q}{%
w\left| a\right| ^{1/w}}\int\limits_{\left| \varepsilon _F\right| }^\infty
\xi ^{1/w}d\xi =-\left| a\right| q\Sigma _q\int\limits_{p_F}^\infty
p^{wq}p^{q-1}dp  \label{ee334}
\end{equation}
Both integrals are divergent, and an upper cutoff momentum is necessary for
their estimation.

\section{Exotic matter with $E>0$ and $P<0$ at $0>w\geq -1$}

Consider formulas (\ref{nn2}) and (\ref{ee2}) when $0>w\geq -1$. This matter
has positive energy density and negative pressure, however, their sum will
be non-negative $P+E\geq 0$. Both integrals (\ref{nn2}) and (\ref{ee2}) are
divergent. However, if we introduce the cutoff value of energy 
\begin{equation}
\varepsilon _0=ap_0^{wq}\ll \varepsilon _F  \label{cut}
\end{equation}
corresponding to the cutoff momentum $p_0\gg p_F$, and change the limits of
integration 
\begin{equation}
\int\limits_{\varepsilon _F}^0...d\varepsilon =\underset{\varepsilon
_0\rightarrow 0}{\lim }\int\limits_{\varepsilon _F}^{\varepsilon
_0}...d\varepsilon  \label{diva}
\end{equation}
\begin{equation}
\int\limits_{p_F}^\infty ...dp=\underset{p_0\rightarrow \infty }{\lim }%
\int\limits_{p_F}^{p_0}...dp  \label{divb}
\end{equation}
then, we obtain finite results that will help us to analyze the behavior of
exotic matter. The divergent terms, substracted from integrals (\ref{nn2})
and (\ref{ee2}), do not depend on temperature and, hence, play no role in
the entropy and heat capacity of exotic matter.

So, the particle number density and the energy density are estimated so

\begin{equation}
n=\frac{\Sigma _q}{wa^{1/w}}\int\limits_{\varepsilon _F}^{\varepsilon
_0}\varepsilon ^{1/w-1}d\varepsilon =\Sigma _q\frac{\varepsilon
_0^{1/w}-\varepsilon _F^{1/w}}{\left| a\right| ^{1/w}}=\Sigma _q\left(
p_0^q-p_F^q\right) =n_0-n_F  \label{nn22}
\end{equation}
where 
\begin{equation}
n_F=\int\limits_0^{p_0}d^qp=\Sigma _qp_F^q  \label{nf}
\end{equation}
and 
\begin{equation}
n_0=\int\limits_0^{p_F}d^qp=\Sigma _qp_0^q  \label{n0}
\end{equation}
The energy density (\ref{ee2}) at $0>w>-1$ is estimated by formula 
\begin{equation}
E=\frac{\Sigma _q}{wa^{1/w}}\int\limits_{\varepsilon _F}^{\varepsilon
_0}\varepsilon ^{1/w}d\varepsilon =\frac{\Sigma _q}{a^{1/w}}\frac{%
\varepsilon _0^{1/w+1}-\varepsilon _F^{1/w+1}}{\left( w+1\right) }=\frac{%
a\Sigma _q}{w+1}\left[ p_0^{\left( w+1\right) q}-p_F^{\left( w+1\right)
q}\right]  \label{en22}
\end{equation}
that is 
\begin{equation}
E=\frac a{\left( w+1\right) \Sigma _q^w}\left[ n_0^{w+1}-\left( n_0-n\right)
^{w+1}\right]  \label{en24}
\end{equation}
Since $p_0\gg p_F$, the particle number density $n=n_0-n_F$ (\ref{en22}) is
approximated by constant value $n\simeq n_0$ (\ref{nn21}), and the energy
density (\ref{en22}) also approximated by a constant value 
\begin{equation}
E\simeq \frac a{\left( w+1\right) \Sigma _q^w}n_0^{w+1}=B=\mathrm{const}
\label{en2g}
\end{equation}
so that the relevant EOS looks like 
\begin{equation}
P\simeq -\left| w\right| B  \label{en2x}
\end{equation}
\begin{equation}
E\simeq B  \label{en2xx}
\end{equation}

According to (\ref{ee2}), the exotic matter with $w=-1$ will have the energy
density 
\begin{equation}
E=a\Sigma _q\mathrm{\ln }\frac{\varepsilon _F}{\varepsilon _0}=aq\Sigma _q%
\mathrm{\ln }\frac{p_0}{p_F}=a\Sigma _q\ln \left( \frac{n_0}{n_F}\right)
\label{en23}
\end{equation}
slightly dependent on $p_F$. Formula (\ref{en23}) bears resemblance with a
logarithmic law in the Hagedorn EOS \cite{Hag} and a logarithmic law in the
transition between $w>-1$ and $w<-1$ in the dark energy \cite{dark3}.
Particularly, taking the energy spectrum (\ref{e-1}) in a 3-dimensional
space, we obtain the EOS 
\begin{equation}
E=\frac{\gamma M^4}{6\pi ^2}\ln \left( \frac{p_0}{p_F}\right)  \label{en93}
\end{equation}

\section{Exotic matter with $E<0$ and $P>0$ at $w\leq -1$}

Consider formulas (\ref{nn3}) and (\ref{ee3}) at $w\leq -1$. This matter has
negative energy density and positive pressure (because $a<0$), however,
their sum will be non-negative $P+E\geq 0$. Integral (\ref{ee3}) is
divergent, and again it is necessary to introduce the cutoff value of energy 
$\varepsilon _0=ap_0^{wq}$ and momentum $p_0\ll p_F$ in order to remove
divergency 
\begin{equation}
\int\limits_\infty ^{\left| \varepsilon _F\right| /T}...dx=\underset{%
\varepsilon _0\rightarrow -\infty }{\lim }\int\limits_{\left| \varepsilon
_0\right| /T}^{\left| \varepsilon _F\right| /T}...dx  \label{div1}
\end{equation}
and 
\begin{equation}
\int\limits_0^{p_F}...dp=\underset{p_0\rightarrow 0}{\lim }%
\int\limits_{p_0}^{p_F}...dp  \label{div2}
\end{equation}

Then, at $w<-1$ the energy density (\ref{ee3}) is determined so 
\begin{equation}
E=-\frac{\Sigma _q}{w+1}\frac{\left| \varepsilon _F\right| ^{1/w+1}-\left|
\varepsilon _0\right| ^{1/w+1}}{\left| a\right| ^{1/w}}=-\frac{\left|
a\right| \Sigma _q}{w+1}\left[ p_F^{\left( w+1\right) q}-p_0^{\left(
w+1\right) q}\right]  \label{ee33}
\end{equation}
that is 
\begin{equation}
E=-\frac{\left| a\right| }{\left( w+1\right) \Sigma _q^w}\left[
n_F^{w+1}-n_0^{w+1}\right] <0  \label{ee34}
\end{equation}
where $n_F$ and $n_0$ are defined in (\ref{nf}) and (\ref{n0}). Since $%
p_0\ll p_F$, the particle number density 
\begin{equation}
n=q\Sigma _q\int\limits_{p_0}^{p_F}p^{q-1}dp=\Sigma _q\left(
p_F^q-p_0^q\right) =n_F-n_0  \label{nn33}
\end{equation}
can be taken in the form $n\simeq n_F$ (\ref{nn3}), and the energy density (%
\ref{ee44}) is estimated so 
\begin{equation}
E\simeq \frac{\left| a\right| \Sigma _q}{w+1}p_0^{\left( w+1\right) q}=%
\mathrm{const}=-B  \label{ee39}
\end{equation}
that results in the following EOS: 
\begin{equation}
E\simeq -B<0\qquad P=\left| w\right| B>0  \label{eos39}
\end{equation}

At $w=-1$ the energy density (\ref{ee3}) is determined so 
\begin{equation}
E=-\left| a\right| \Sigma _q\ln \frac{\varepsilon _0}{\varepsilon _F}%
=-\left| a\right| q\Sigma _q\ln \frac{p_F}{p_0}  \label{ee4}
\end{equation}
that is 
\begin{equation}
E=-\left| a\right| \Sigma _q\ln \frac{n_F}{n_0}\simeq -\left| a\right|
\Sigma _q\ln \frac n{n_0}  \label{ee44}
\end{equation}
because $n\simeq n_F\gg n_0$. This formula also bears resemblance with
logarithmic laws in the Hagedorn EOS \cite{Hag} and transition between $w>-1$
and $w<-1$ in the dark energy \cite{dark}.

\section{Low temperature expansion}

Formulas (\ref{nno})-(\ref{eew}) can be presented in the universal form

\begin{equation}
n=\frac{\Sigma _q}{\left| w\right| }\frac{T^{1/w}}{\left| a\right| ^{1/w}}%
J\left( \lambda \right)  \label{i0}
\end{equation}
and 
\begin{equation}
E=\mathrm{sign}\left( a\right) \frac{\Sigma _q}{\left| w\right| }\frac{%
T^{1/w+1}}{\left| a\right| ^{1/w}}J\left( \lambda \right)  \label{ie}
\end{equation}
where integral

\begin{equation}
J\left( \lambda \right) =\int\limits_0^\infty g\left( x\right) f_p\left(
x,\lambda \right) dx  \label{i}
\end{equation}
includes the distribution function $f_p$ (\ref{ff}) and function 
\begin{equation}
g\left( x\right) =x^{1/w-1}  \label{g1}
\end{equation}
or 
\begin{equation}
g\left( x\right) =x^{1/w}  \label{g2}
\end{equation}
corresponding to the particle number density and energy density,
respectively.

Integrating by parts, we have 
\begin{equation}
J\left( \lambda \right) =\left. G\left( x\right) f_p\left( x,\lambda \right)
\right| _0^\infty -\int\limits_0^\infty G\left( x\right) f_p^{\prime }\left(
x,\lambda \right) dx  \label{i1}
\end{equation}
where 
\begin{equation}
f_p^{\prime }\left( x,\lambda \right) =\frac{\partial f_p\left( x,\lambda
\right) }{\partial x}=-\frac{\mathrm{sign}\left( a\right) \exp \left[ 
\mathrm{sign}\left( a\right) \left( x-\lambda \right) \right] }{\left\{ \exp
\left[ \mathrm{sign}\left( a\right) \left( x-\lambda \right) \right]
+1\right\} ^2}=-\frac{\mathrm{sign}\left( a\right) \exp \left( x-\lambda
\right) }{\left( \exp \left( x-\lambda \right) +1\right) ^2}  \label{fpr}
\end{equation}
and 
\begin{equation}
G\left( x\right) =\int g\left( x\right) dx  \label{g}
\end{equation}

Expression 
\begin{equation}
J_0=\left. G\left( x\right) f_p\left( x,\lambda \right) \right| _0^\infty =%
\underset{x\rightarrow \infty }{\lim }\left[ G\left( x\right) f_p\left(
x,\lambda \right) \right] -\underset{x\rightarrow 0}{\lim }\left[ G\left(
x\right) f_p\left( x,\lambda \right) \right]  \label{G}
\end{equation}
in the light of (\ref{ff2}) and (\ref{ff3}), is simplified so 
\begin{equation}
J_0=-\underset{x\rightarrow 0}{\lim }G\left( x\right) \qquad a>0
\label{G1}
\end{equation}
\begin{equation}
J_0=\underset{x\rightarrow \infty }{\lim }G\left( x\right) \qquad a<0
\label{G2}
\end{equation}
If quantity $J_0$ is divergent, then, we introduce some finite cutoff value 
\begin{equation}
x_0\ll \lambda \qquad a>0  \label{ct1}
\end{equation}
or 
\begin{equation}
x_0\gg \lambda \qquad a<0  \label{ct2}
\end{equation}
So, we can present $J_0$ (\ref{G1})-(\ref{G2}) in a universal form 
\begin{equation}
J_0=-\mathrm{sign}\left( a\right) G\left( x_0\right)  \label{ct}
\end{equation}
Therefore, integral (\ref{i1}) is immediately written in the form 
\begin{equation}
J\left( \lambda \right) =-\mathrm{sign}\left( a\right) G\left( x_0\right)
-\int\limits_0^\infty G\left( x\right) f_p^{\prime }\left( x,\lambda \right)
dx  \label{i11}
\end{equation}

Let us expand function $G\left( x\right) $ in the Taylor series \cite{Ziman} 
\begin{equation}
G\left( x\right) =G\left( \lambda \right) +\sum\limits_{k=1}^{k=\infty }%
\frac{g^{\left( k-1\right) }\left( \lambda \right) }{k!}\left( x-\lambda
\right) ^k  \label{tay}
\end{equation}
where 
\begin{equation}
g^{\left( k\right) }\left( x\right) =\frac{\partial ^kg\left( x\right) }{%
\partial x^k}  \label{gk}
\end{equation}

Substituting (\ref{tay}) in (\ref{i11}) we have 
\begin{equation}
J\left( \lambda \right) =-\mathrm{sign}\left( a\right) G\left( x_0\right)
-G\left( \lambda \right) \int\limits_0^\infty f_p^{\prime }\left( x,\lambda
\right) dx-\sum\limits_{k=1}^{k=\infty }\frac{g^{\left( k-1\right) }\left(
\lambda \right) }{k!}\int\limits_0^\infty \left( x-\lambda \right)
^kf_p^{\prime }\left( x,\lambda \right) dx  \label{i2}
\end{equation}
In the light of (\ref{ff})-(\ref{ff3}), the first term in (\ref{i2}) is
simplified so 
\begin{equation}
-G\left( \lambda \right) \int\limits_0^\infty f_p^{\prime }\left( x,\lambda
\right) dx=-\left. G\left( \lambda \right) f_p\left( x,\lambda \right)
\right| _0^\infty =\mathrm{sign}\left( a\right) G\left( \lambda \right)
\label{i22}
\end{equation}
Hence 
\begin{equation}
J\left( \lambda \right) =-\mathrm{sign}\left( a\right) G\left( x_0\right) +%
\mathrm{sign}\left( a\right) G\left( \lambda \right)
-\sum\limits_{k=1}^{k=\infty }\frac{g^{\left( k-1\right) }\left( \lambda
\right) }{k!}\int\limits_0^\infty \left( x-\lambda \right) ^kf_p^{\prime
}\left( x,\lambda \right) dx  \label{i3}
\end{equation}
where $f_p^{\prime }\left( x,\lambda \right) $ is determined by (\ref{fpr}).
At low temperature ($\lambda =\mu /T\gg 1$) integral (\ref{i3}) is
approximated by formula 
\begin{equation}
J\left( \lambda \right) \cong \mathrm{sign}\left( a\right) \left[ G\left(
\lambda \right) -G\left( x_0\right) +\sum\limits_{k=1}^{k=\infty }g^{\left(
k\right) }\left( \lambda \right) C_k\right]  \label{i4}
\end{equation}
with coefficients 
\begin{eqnarray}
C_k=\frac 1{k!}\int\limits_0^\infty \left( x-\lambda \right) ^{2k}\frac{\exp
\left[ \mathrm{sign}\left( a\right) \left( x-\lambda \right) \right] }{%
\left\{ \exp \left[ \mathrm{sign}\left( a\right) \left( x-\lambda \right)
\right] +1\right\} ^2}dx=  \nonumber \\
\quad =\int\limits_{-\lambda }^\infty x^{2k}\frac{\exp \left( x\right) }{%
\left( \exp \left( x\right) +1\right) ^2}dx\cong \int\limits_{-\infty
}^\infty x^k\frac{\exp \left( x\right) }{\left( \exp \left( x\right)
+1\right) ^2}dx &&  \label{co}
\end{eqnarray}

Note that all odd coefficients (\ref{co}) tend to zero 
\begin{equation}
C_{2k+1}\rightarrow 0  \label{co1}
\end{equation}
Integral (\ref{i4}) can be written in explicit form 
\begin{equation}
J\left( \lambda \right) =\mathrm{sign}\left( a\right) \left[ G\left( \lambda
\right) -G\left( x_0\right) +g^{\prime }\left( \lambda \right) \frac{\pi ^2}%
6+g^{\prime \prime \prime }\left( \lambda \right) \frac{7\pi ^4}{360}%
+...\right]  \label{i55}
\end{equation}
that is 
\begin{equation}
J\left( \lambda \right) =G\left( \lambda \right) -\underset{x_0\rightarrow
0}{\lim }G\left( x_0\right) +g^{\prime }\left( \lambda \right) \frac{\pi ^2}%
6+g^{\prime \prime \prime }\left( \lambda \right) \frac{7\pi ^4}{360}%
+...\qquad a>0  \label{i5}
\end{equation}
and 
\begin{equation}
J\left( \lambda \right) =\underset{x_0\rightarrow \infty }{\lim }G\left(
x_0\right) -G\left( \lambda \right) -g^{\prime }\left( \lambda \right) \frac{%
\pi ^2}6-g^{\prime \prime \prime }\left( \lambda \right) \frac{7\pi ^4}{360}%
+...\qquad a<0  \label{i6}
\end{equation}
For arbitrary function $g\left( x\right) $ formula (\ref{i55}) determines a
low temperature expansion of the relevant thermodynamical quantity.

\section{Fermi level at low temperature}

According to (\ref{g2}) and (\ref{g}), we have\textrm{\ } 
\begin{equation}
G\left( x\right) =wx^{1/w}  \label{g3}
\end{equation}
Substituting function (\ref{g3}) in integrals (\ref{i55}), we obtain

\begin{equation}
J\left( \lambda \right) =\mathrm{sign}\left( a\right) \left( w\lambda
^{1/w}-wx_0^{1/w}+\frac{1-w}w\frac{\pi ^2}6\lambda ^{1/w-2}\right)
\label{i9}
\end{equation}
where the cutoff value $x_0$ is taken according to (\ref{ct1}) and (\ref{ct2}%
).

Substituting (\ref{i9}) in (\ref{i0}) we get the particle number density 
\begin{equation}
n=\mathrm{sign}\left( a\right) \mathrm{sign}\left( w\right) \left(
n_F-n_0\right)  \label{nz1}
\end{equation}
where

\begin{equation}
n_F=\Sigma _q\frac{\left| \mu \right| ^{1/w}}{\left| a\right| ^{1/w}}\left(
1+\frac{1-w}{w^2}\frac{\pi ^2}6\frac{T^2}{\mu ^2}\right)  \label{nz2}
\end{equation}
and 
\begin{equation}
n_0=\Sigma _q\frac{\left| \mu _0\right| ^{1/w}}{\left| a\right| ^{1/w}}%
=\Sigma _qp_0^q=\mathrm{const}  \label{nz3}
\end{equation}

At zero temperature $\mu \rightarrow \varepsilon _F$, and formula (\ref{nz2}%
) is reduced to \textrm{\ } 
\begin{equation}
n_F=\Sigma _q\frac{\left| \varepsilon _F\right| ^{1/w}}{a^{1/w}}=\Sigma
_qp_F^q  \label{nz4}
\end{equation}
that coincides with (\ref{nf}).\textrm{\ }Hence, the Fermi energy level at
low temperature is approximated by formula 
\begin{equation}
\left| \mu \right| \cong \left| \varepsilon _F\right| \left( 1-\frac{1-w}w%
\frac{\pi ^2}6\frac{T^2}{\varepsilon _F^2}\right)  \label{nz5}
\end{equation}
Note that it does not depend on the sign of $a$, neither divergency of play (%
\ref{i9}) is reflected here.

For nonrelativistic EOS with $w=2/3$\ we get a well known expression \cite
{Ziman} 
\begin{equation}
\mu \cong \varepsilon _F\left( 1-\frac{\pi ^2}{12}\frac{T^2}{\varepsilon _F^2%
}\right)  \label{nz6}
\end{equation}
At $w<0$ the absolute value of Fermi level $\left| \mu \right| $ always
exceeds the same at zero temperature $\left| \varepsilon _F\right| $,
particularly, at $w=-1$ the low-temperature approximation of the Fermi level
is the following 
\begin{equation}
\left| \mu \right| \cong \left| \varepsilon _F\right| \left( 1+\frac{\pi ^2}3%
\frac{T^2}{\varepsilon _F^2}\right)  \label{nz7}
\end{equation}

\section{Energy density at low temperature}

According to (\ref{g2}) and (\ref{g}), we have 
\begin{equation}
G\left( x\right) =\frac w{w+1}x^{1/w+1}\qquad w\neq -1  \label{g4}
\end{equation}
and 
\begin{equation}
G\left( x\right) =\ln x\qquad w=-1  \label{g5}
\end{equation}

Substituting (\ref{g4})-(\ref{g5}) in (\ref{i55}), we obtain 
\begin{equation}
\mathrm{sign}\left( a\right) J\left( \lambda \right) =\frac w{w+1}\lambda
^{1/w+1}-\frac w{w+1}x_0^{1/w+1}+\frac{\lambda ^{1/w-1}}w\frac{\pi ^2}6+%
\frac{\left( 1-w\right) \left( 1-2w\right) }{w^3}\frac{7\pi ^4}{360}\lambda
^{1/w-3}  \label{ie1}
\end{equation}
when $w\neq -1$, while 
\begin{equation}
\mathrm{sign}\left( a\right) J\left( \lambda \right) =\ln \frac \lambda
{x_0}-\frac{\pi ^2}6\frac 1{\lambda ^2}-\frac{7\pi ^4}{60}\frac 1{\lambda ^4}
\label{ie2}
\end{equation}
when $w=-1$. The cutoff value $x_0=\varepsilon _0/T$ is taken from
conditions (\ref{ct1}) and (\ref{ct2}) in accordance to the sign of $a$.

Substituting (\ref{ie1}) in (\ref{ie}) we find the energy density\textrm{\ } 
\begin{equation}
E=\frac{\mathrm{sign}\left( w\right) }{w+1}\frac{\Sigma _q}{\left| a\right|
^{1/w}}\left\{ \left| \mu \right| ^{1/w+1}\left[ 1+\frac{w+1}{w^2}\frac{\pi
^2}6\frac{T^2}{\mu ^2}+\frac{\left( 1-w^2\right) \left( 1-2w\right) }{w^4}%
\frac{7\pi ^4}{360}\frac{T^4}{\mu ^4}\right] -\left| \varepsilon _0\right|
^{1/w+1}\right\}  \label{ie11}
\end{equation}
when $w\neq -1$. Substituting (\ref{ie2}) in (\ref{ie}) we find the energy
density\textrm{\ } 
\begin{equation}
E=\frac{\Sigma _q}{\left| a\right| ^{1/w}}\left[ \ln \frac{\left| \mu
\right| }{\left| \varepsilon _0\right| }-\frac{\pi ^2}6\frac{T^2}{\mu ^2}-%
\frac{7\pi ^4}{60}\frac{T^4}{\mu ^4}\right]  \label{ie22}
\end{equation}
when $w=-1$. We can check formula (\ref{ie11}) for regular ultrarelativistic
matter with EOS $P=E/3$ and in 3-dimensional space \cite{qcd97}\textrm{\ } 
\begin{equation}
P=\frac E3=\frac \gamma {24\pi ^2}\left( \mu ^4+2\pi ^2T^2\mu ^2+\frac{7\pi
^4}{15}T^4\right)  \label{i23}
\end{equation}

Substituting (\ref{nz5}) in (\ref{ie11}), we obtain a low-temperature
expansion of the energy density at $w\neq -1$: \textrm{\ } 
\begin{equation}
E=\frac{\mathrm{sign}\left( w\right) }{w+1}\frac{\Sigma _q}{\left| a\right|
^{1/w}}\left[ \left| \varepsilon _F\right| ^{1/w+1}\left( 1+\frac{w+1}w\frac{%
\pi ^2}6\frac{T^2}{\left| \varepsilon _F\right| ^2}\right) -\left|
\varepsilon _0\right| ^{1/w+1}\right]  \label{ie3}
\end{equation}
Formula (\ref{ie3}) can rewritten so 
\begin{equation}
E=E_0+\frac{\pi ^2}6\Sigma _q\frac{\left| \varepsilon _F\right| ^{1/w+1}}{%
\left| w\right| \left| a\right| ^{1/w}}\frac{T^2}{\left| \varepsilon
_F\right| ^2}  \label{ie4}
\end{equation}
where 
\begin{equation}
E_0=\frac{\mathrm{sign}\left( w\right) }{w+1}\frac{\Sigma _q}{\left|
a\right| ^{1/w}}\left( \left| \varepsilon _F\right| ^{1/w+1}-\left|
\varepsilon _0\right| ^{1/w+1}\right) =\frac{\mathrm{sign}\left( w\right) }{%
\left( w+1\right) }\frac{\left| a\right| }{\Sigma _q^w}\left(
n_F^{w+1}-n_0^{w+1}\right)  \label{ie44}
\end{equation}
is the energy density at zero temperature. Expression (\ref{ie44}) embraces
formulas (\ref{ee1}), (\ref{ee2}), (\ref{ee3}) and (\ref{ee4}) when $w\neq
-1 $.

Substituting (\ref{nz5}) in (\ref{ie22}), we obtain a low-temperature
expansion of the energy density at $w=-1$: 
\begin{equation}
E=E_0+\frac{\pi ^2}6\Sigma _q\left| a\right| \frac{T^2}{\left| \varepsilon
_F\right| ^2}  \label{ie5}
\end{equation}
where 
\begin{equation}
E_0=\Sigma _q\left| a\right| \ln \frac{\left| \varepsilon _F\right| }{\left|
\varepsilon _0\right| }=\Sigma _q\left| a\right| \ln \frac{n_0}{n_F}
\label{ie55}
\end{equation}
is the energy density at zero temperature, that coincides with (\ref{en23})
and (\ref{ee4}).

\section{Entropy and heat capacity at low temperature}

Expressions (\ref{ie4}) and (\ref{ie5}) allow to obtain the entropy density $%
S$ and the heat capacity $C_V$ according to standard formulas \cite{LL5} 
\begin{equation}
S=-\frac{\partial \left( T\ln Z\right) }{\partial T}=\frac{\partial P}{%
\partial T}\qquad C_V=T\frac{\partial S}{\partial T}  \label{ent}
\end{equation}
where the pressure is $P=wE$ (\ref{xi}). Substituting (\ref{ie4}) and (\ref
{ie5}) in (\ref{ent}), we find general formula 
\begin{equation}
S=C_V=\mathrm{sign}\left( w\right) \frac{\pi ^2}3\Sigma _q\frac{T\left|
\varepsilon _F\right| ^{1/w-1}}{\left| a\right| ^{1/w}}=\mathrm{sign}\left(
w\right) \frac{\pi ^2}3\Sigma _q\frac T{\left| \varepsilon _F\right| }p_F^q=%
\mathrm{sign}\left( w\right) \frac{\pi ^2}3\frac T{\left| \varepsilon
_F\right| }n_F  \label{ent2}
\end{equation}
which is valid for any $w$. The density $n_F$ is defined by (\ref{nz4}).
Particularly, at $w=1$ the heat capacity is 
\begin{equation}
S=C_V=\frac{\pi ^2}3\Sigma _q\frac T{\left| a\right| }  \label{ent1}
\end{equation}
that after parametrization $a=m^2$ yields the heat capacity of 'absolute
stiff' matter \cite{TV2011e}. The regular nonrelativistic Fermi gas with the
energy spectrum $\varepsilon _p=\frac{p^2}/(2m)$ and EOS $P=2/3E$ yields has
the heat capacity \cite{Ziman} 
\begin{equation}
S=C_V=\frac{\pi ^2}3\Sigma _qm^{3/2}\sqrt{\left| \varepsilon _F\right| }T
\label{ent3}
\end{equation}

The sign of the energy density $E$ is not sufficient, but the sign of $w$
play the main role, and exotic fermion matter with $w<0$ will have negative
entropy and negative heat capacity, particularly 
\begin{equation}
S=C_V=-\frac{\pi ^2}3\Sigma _q\frac{\left| a\right| T}{\left| \varepsilon
_F\right| ^2}  \label{ent4}
\end{equation}
at $w=-1$.

\section{Conclusion}

For and arbitrary relation between the pressure and energy density $P=wE$ (%
\ref{xi}) at constant $w$, the EOS can be modeled by and ideal gas of free
quasi-particles with the universal energy spectrum $\varepsilon _p=ap^{wq}$ (%
\ref{e0}). The sign of $w$ and $a$ can be arbitrary, and thermodynamical
functions of this gas are determined by formulas (\ref{nno}) and (\ref{eeo}%
). We have analyzed in detail a Fermi gas of such quasi-particles at zero
temperature.

For regular matter with positive pressure $P>0$ and positive energy density $%
E>0$ (correspond to $w>0$ and $a>0$), the particle number density $n$ and
energy density $E$ are given by formulas (\ref{nn1}), (\ref{ee1}), and (\ref
{ee11}).

For exotic matter with negative pressure $P<0$ and positive energy density $%
E>0$ ($w<0$ and $a>0$), the particle number density and energy density are
given by formulas (\ref{nn2}) and (\ref{ee2}). At $w<-1$, when $P+E<0$, the
energy density is given by formula (\ref{ee22}) which characterizes the
regular matter (\ref{ee1}) and reveals proportionality 
\begin{equation}
E\sim n^{w+1}  \label{law}
\end{equation}
At $w\geq -1$ the energy density is divergent, and a cutoff momentum $p_0$
is introduced for its estimation. At $w=-1$ the energy density is given by
formula (\ref{ee23}) and at $0>w>-1$ (when $P+E>0$) the energy density
attains constant value (\ref{en2g}) which depends on $p_0$. However, the law
(\ref{law}) is not working now, but the matter admits description in the
frames of the 'bag' model $P=-E=\mathrm{const}$.

For exotic matter with positive pressure $P>0$ and negative energy density $%
E<0$ ($w>0$ and $a<0$), the particle number density and energy density are
given by formulas (\ref{nn3}) and (\ref{ee3}). At $0>w>-1$ (when $P+E<0$),
the energy density is given by formula (\ref{p1x}) which obeys the law (\ref
{law}).

At $w\leq -1$ the energy density is divergent, and a cutoff momentum $p_0$
is introduced again for its estimation. At $w=-1$ the energy density is
given by formula (\ref{ee44}), while at $w<-1$ (when $P+E>0$), the energy
density attains constant value (\ref{ee39}). Again, proportionality (\ref
{law}) is not valid now.

In order to estimate the thermodynamical functions of Fermi gas at finite
temperature, it is necessary to develop formulas (\ref{i}), (\ref{i}), (\ref
{i5}), (\ref{i6}). The particle number density $n$ at low temperature is
determined by expressions (\ref{nz1})-(\ref{nz3}). The Fermi level $\mu $ at
low temperature is determined by formula (\ref{nz5}). Its peculiar property
is that it exceeds the Fermi level at zero temperature $\varepsilon _F$ when 
$1>w>0$, meanwhile $\mu <\varepsilon _F$ when $w<0$ or when $w>1$. A
low-temperature expansion of the energy density is determined by formula (%
\ref{ie3}) at $w\neq -1$, and by formula (\ref{ie5}) at $w=-1$. The entropy
density and heat capacity at low temperature is given by the single formula (%
\ref{ent2}). It is linear dependent on temperature. However, exotic matter
with negative $w<0$ has negative entropy and negative heat capacity, see,
for example, formula (\ref{ent4}) at $w=-1$.

The theory of ideal Fermi gas of quasi-particles in 3-dimensional space
(whose energy spectrum is $\varepsilon _p=ap^{3w}$) can be immediately
applied to compact stellar objects containing hypothetical material with
arbitrary EOS $P=wE$, where $w$ is constant. However, we should not forget
that the central pressure must be positive. Only the regular matter ($a>0$
and $w>0$) and exotic matter with positive pressure and negative energy
density (when $a<0$ and $w<0$) are suitable for stable stellar models.
Calculations are in progress.

For the EOS $P=-E$ the energy density and particle number density at zero
temperature have a logarithmic link (\ref{ie55}) similar to the Hagedorn
EOS. This also deserves more analysis.

The authors are grateful to Erwin Schmidt for discussions.

\newpage


\end{document}